\documentclass[11pt]{article}
\usepackage[letterpaper,margin=1in]{geometry}
\usepackage{fontspec}
\usepackage{amsmath,amssymb,graphicx,caption,cite}
\usepackage[hidelinks,pdfversion=1.7]{hyperref}
\usepackage{microtype}
\usepackage{xcolor}
\graphicspath{{figures/}}
\hypersetup{pdftitle={A 28nm 27,648-Spin Multichip Digital Ising Accelerator with Pegasus Connectivity}}
\hypersetup{pdfauthor={Tong Wu et al.}}
\begin{document}
\noindent{\large\bfseries A 28nm 27,648-Spin Multichip Digital Ising Accelerator with Pegasus Connectivity}
\begin{center}
Tong Wu\textsuperscript{1}, Atharva Raut\textsuperscript{1}, Ian Khor\textsuperscript{2}, Ziyad Alswaidan\textsuperscript{1}, Ting-Yu Lee\textsuperscript{1},\\
Hongchang Kuang\textsuperscript{1}, Navid Anjum Aadit\textsuperscript{3}, Anand Raju\textsuperscript{1}, Dhruv Chinmay\textsuperscript{1},\\
Siddharth Das\textsuperscript{1}, Ken Mai\textsuperscript{1}, Kerem Camsari\textsuperscript{4}, Tathagata Srimani\textsuperscript{1}\\[0.4em]
\textsuperscript{1}Carnegie Mellon University; \textsuperscript{2}University of Michigan;\\
\textsuperscript{3}Stanford University; \textsuperscript{4}University of California, Santa Barbara
\end{center}
\begin{abstract}
We report a 28 nm four-chip Ising accelerator with 27,648 spins, degree-15 Pegasus connectivity, and 10b coefficients. Boundary-state streaming overlaps interchip transfers with spin updates, achieving 98.03\% simulated weak-scaling efficiency. At 140 MHz, the four-chip system delivers 30.24G peak updates/s at 1.2 pJ/update including I/O power, with 11.5$\times$ the throughput and approximately one-quarter the logic-plus-SRAM area per spin of a prior 28 nm multichip design. We demonstrate MaxCut, spin glass, frustrated loops, factorization, and 3SAT.
\end{abstract}

Ising accelerators address combinatorial optimization problems (COPs) by updating coupled binary variables, or spins, to find low-energy configurations representing solutions \cite{lucas,yamaoka}. Scaling to larger problems is limited by coefficient storage and on-chip/multi-chip interconnect. Fully connected implementations support arbitrary pairwise interactions but require $O(N^2)$ coefficients for $N$ spins \cite{statica,kawamura}. Dense-to-sparse transformations reduce connectivity by introducing additional spins \cite{dense_sparse}. Minor graph embedding, which uses coupled physical spins to realize connections absent from a fixed hardware graph, can further increase the spin count~\cite{choi,minorminer}. These transformations trade connectivity for additional spins, motivating multichip integration for larger problems.

While multichip integration increases capacity, exchanging spin states can still stall dependent updates. Prior two- and nine-chip implementations use King's graphs with at most eight neighbors per spin (degree 8) \cite{takemoto2020,takemoto2021}, increasing embedding overhead for dense problems. Fully connected hardware reduces transfer overhead through compressed spin transfers and double buffering \cite{kawamura}, but communication still contributes to system energy and throughput overhead. A reconfigurable 8K-spin design supports up to 31 two- and three-body interactions \cite{kim_sim}, but provides different connectivity within and across chips. The challenge is therefore to maintain high connectivity across chip boundaries while sustaining high update throughput and low energy per attempted spin update.

We address this challenge with a 28 nm four-chip Ising accelerator that implements a 27,648-spin Pegasus graph with the same connection density within and across chips (Fig.~1). The accelerator achieves higher spin-update throughput and lower energy per attempted update vs. prior multi-chip designs. Pegasus supports up to 15 neighbors per spin, compared with six for Chimera and eight for King's graphs \cite{pegasus,niazi,bunyk,bae}. For the illustrated 32-spin fully connected problem, embedding requires 133 Pegasus spins versus 1,885 King's-graph spins, a 14.2$\times$ reduction. To sustain updates at this connectivity, banked static random-access memory (SRAM) supplies coefficients to parallel spin-update circuits. Boundary-state streaming overlaps interchip neighbor-state transfers with independent spin updates. Together, these techniques provide 6,912 spins per chip with 10b couplings and biases and achieve 98.03\% simulated weak-scaling efficiency, defined as four-chip spin-update throughput divided by four times the single-chip throughput.

Fig.~2 shows our chip architecture: nine Ising units (IUs), coefficient SRAM, spin registers, and chip-to-chip (C2C) interfaces. Each IU contains six processing elements (PEs), providing 54 parallel spin updates per chip per cycle. The PEs are time-multiplexed across 32 subgraphs, each containing nine 24-spin Pegasus cells (Fig.~3). Spins are partitioned into four colors, without directly coupled spins being in the same color \cite{niazi}. Spins in one color update each cycle using the stored spins in the other three colors. The four cycles update a 216-spin subgraph; traversing the $8\times4$ subgraph array completes one Monte Carlo sweep (MCS), updating all 6,912 spins in 128 cycles without stalls. A shared 7b address, comprising a 5b subgraph index and a 2b color index, controls coefficient reads, neighbor selection, and spin-register writeback.

To supply all six PEs in an IU, four $128\times240$b single-port SRAM banks deliver 960b coefficient data each cycle. Each PE receives fifteen signed 10b couplings and one signed 10b bias; the nine IUs contain 36 banks with a total memory capacity of 1.1 Mb. Reading the complete 160b coefficient record for each spin in one access allows all 15 interactions to be evaluated concurrently. All coefficients are stored in SRAM during updates, including those for interchip interactions. Fixed Pegasus connections select neighbor registers without storing an address alongside each coefficient. Coefficients are zeroed to disable unneeded interactions. Reprogramming a problem therefore changes neither the memory-access sequence nor the coefficient bandwidth per update.

Each PE implements probabilistic updates \cite{aadit,camsari} for the Ising energy $H=-\sum_{i<j}J_{ij}s_i s_j-\sum_i h_i s_i$, where $s_i\in\{-1,+1\}$ is a spin, $J_{ij}$ a coupling, and $h_i$ a bias. Sign-selection logic chooses $J_{ij}$ or $-J_{ij}$ according to $s_j$, and an adder tree sums the interactions and bias to form the local field $I_i$. A hyperbolic-tangent lookup table (LUT) and comparator implement $s_i=\operatorname{sgn}[\tanh(\beta I_i)-r_i]$, where $\beta$ is inverse temperature and $r_i$ is uniform pseudorandom noise on $[-1,1]$. The clamped field's 7b magnitude addresses the LUT; sign logic reconstructs its output, and a linear-feedback shift register (LFSR) supplies the random input. Scaling the coefficients by inverse temperature ($\beta$) during programming eliminates the need for annealing multipliers in each PE.

We extend this datapath across chips by storing received states in boundary registers that feed the same neighbor-selection circuits as on-chip states. Since coefficients remain in local SRAM, only 1b spin states cross the C2C interface. The states of three edge IUs are combined into 42b north-edge or 54b south-edge packets (Fig.~2), transmitted as 6b or 7b words over eight input/output (I/O) cycles with padding. Fixed bit order selects the destination register without per-spin address headers. Sending all required states, including unchanged states, makes the transfer sequence independent of spin activity.

To hide transfer latency, we order the subgraphs so that a boundary update precedes the receiving chip's dependent update by half a sweep (Fig.~3). The PEs process intervening subgraphs while the C2C interface transfers the boundary states. Four cycles per subgraph give a 64-compute-cycle interval for the boundary exchange. The receiver intervals of the north- and south-edge alternate, allowing a shared output interface to serve both neighbors and reuse output pads. Each transmitter forwards a source-synchronous I/O clock divided from its compute clock; the divider is selected to complete transfers within the available compute interval. Asynchronous first-in, first-out (FIFO) buffers connect the compute and I/O clock domains. Synchronized Gray-coded pointers support full/empty detection, while ready/valid handshaking controls transfers. If required input data are unavailable or an output FIFO is full, the controller holds the subgraph/color address and suppresses spin writeback. When packets arrive before the dependent update, computation proceeds without communication stalls.

With this overlap, each added chip contributes 6,912 spins and its update throughput. At a common compute-clock frequency, the ideal throughput for $C$ chips is $CR_1$, where $R_1$ is the single-chip rate. Scaling efficiency is defined as $R_C/(CR_1)$, where $R_C$ is aggregate throughput at fixed work per chip; simulation for more than two chips gives 98.03\% (Fig.~6). The local address width, packet format, and per-chip coefficient bandwidth remain unchanged as chips are added, supporting unbounded streaming extension subject to clock distribution and link timing. Package-level clock distribution limits our four-chip prototype to 140 MHz, while a single chip can operate at 250 MHz, 75.1 mW. The four chips update one coupled graph, with boundary registers supplying neighboring chips' states to the local datapaths.

We demonstrate four-chip operation with planted Ising and spin-glass problems (Fig.~4). For planted Ising, we consider two different instances of MaxCut problems - (a) P2 (Fig.~4), where successive spin maps recover the programmed image (the text 'ISSCC') in 5.33 $\mu$s, and (b) P1 (Fig.~4), a 27,648 spin MaxCut which maximizes the weight of edges between two graph partitions that reaches the exact ground-state energy $-13{,}824$ at a $\beta=8$ quench within 156 clock cycles (corresponding to 1.11 $\mu$s @ 140 MHz). Separately, for a 27,648-spin glass with random signed couplings, 500 samples at $\beta=4$ concentrate at lower energies vs. a CPU baseline~\cite{metropolis}. Annealing~\cite{kirkpatrick} from $\beta=0.5$ to 4 gives the convergence times of 1.404 ms for our accelerator (compared to 1.45 s for the CPU). 

Additionally, to evaluate competing interactions, we program frustrated-loop problems, in which not all coupling preferences can be satisfied (Fig.~5)~\cite{hen}. The 13.6k-spin study varies loop density $\alpha$, the number of planted loops per spin, from 0.1 to 4. We iteratively search for the best annealing schedule per $\alpha$ and evaluate time to solution at 99\% success probability ($\mathrm{TTS}_{99}$)~\cite{ronnow}, with an energy within 2\% of the reference minimum as the success criterion. A separate 27,648-spin test at $\alpha=8$ reaches the exact ground state in two of sixteen runs, at 0.207 s and 0.636 s. 

We also demonstrate factorization~\cite{camsari} and three-literal Boolean satisfiability (3SAT) on the same datapath through quadratic unconstrained binary optimization (QUBO), which expresses constraints as a quadratic objective over binary variables \cite{lucas}. The factorization experiment recovers $899=31\times29$, using 205 logical spins mapped onto 2,601 spins on the accelerator (Fig.~5). For 3SAT, each clause is an OR of three variables or their complements. The 3SAT-derived Ising problems span 200 to 2,000 clauses and 0.9k to 9.2k spins, with 2.21--3.66 $\mu$s of $\mathrm{TTS}_{99}$ using an energy gap of at most 2\% as the success criterion. 

Fig.~6 compares our chip with prior digital Ising accelerators \cite{su,kawamura,kim_sim,mzephyr,cobi}.  With 10b coefficients, our chips deliver 30.24G peak updates/s at 140 MHz, 11.5$\times$ the throughput of a recent 28 nm multichip implementation \cite{kim_sim}. At 0.7 V core and 0.9 V I/O supplies, we measure our energy at 1.2 pJ/spin-update including I/O power. Core-only energy is 0.939 pJ/update, 78.1$\times$ lower vs. prior multi-chip design \cite{kim_sim}. The logic and SRAM area is 74.1 $\mu$m$^2$/spin, approximately one-quarter of that reported in \cite{kim_sim}. Throughput counts spin-update attempts, including those that retain the previous state.

Fig.~7 shows our 28 nm chip and four-chip test board. Each die occupies 1.5 mm$^2$, with a 0.846 mm$^2$ core. The functional SRAM-plus-logic area is 0.512 mm$^2$. A host Xilinx ZCU104 FPGA loads coefficients and random seeds and captures spin states. We demonstrate five optimization workloads on one programmable Ising datapath. Banked coefficient storage sustains parallel spin updates, while overlapped boundary transfers extend the graph across chips.

\newcommand{\paperfigure}[3]{%
  \clearpage
  \begin{figure}[!ht]
    \centering
    \includegraphics[width=\textwidth,height=0.82\textheight,keepaspectratio]{#1.pdf}
    \caption{#2}\label{#3}
  \end{figure}
}
\paperfigure{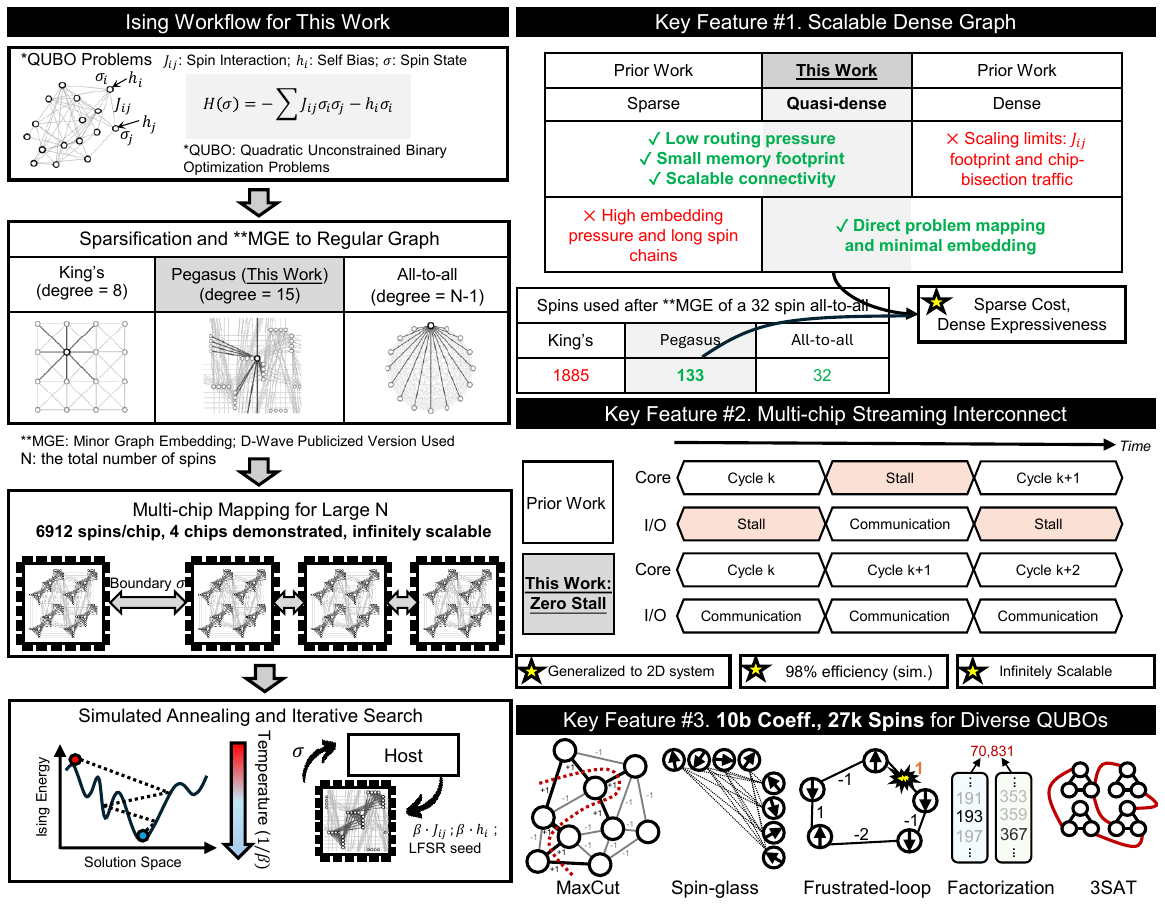}
{Ising/QUBO workflow and key features: degree-15 Pegasus connectivity, reduced embedding overhead, and boundary-state streaming across four chips.}
{fig:overview}
\paperfigure{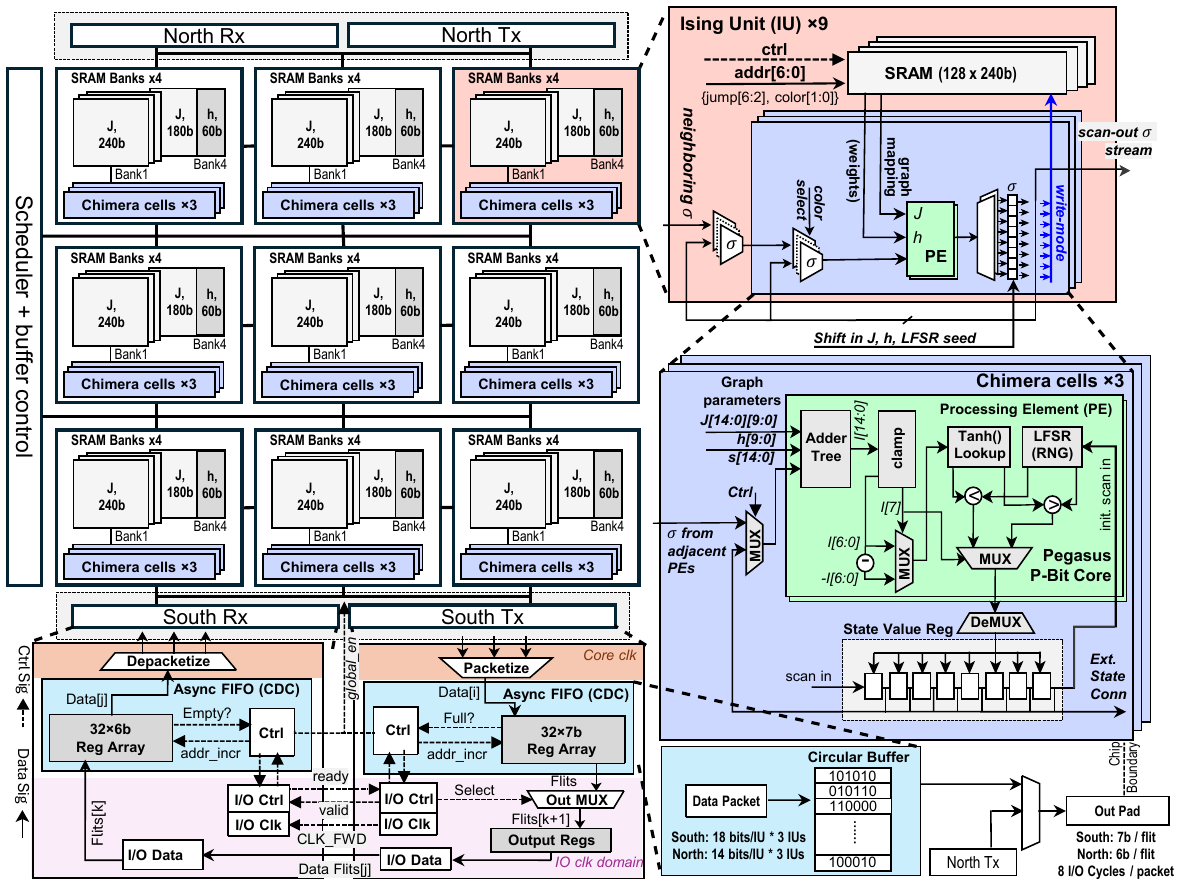}
{Chip architecture, banked coefficient SRAM, parallel spin-update datapath, and source-synchronous C2C interface. FIFO buffers decouple compute and I/O clocks. Tx/Rx: transmit/receive; CDC: clock-domain crossing; RNG: random-number generator.}
{fig:architecture}
\paperfigure{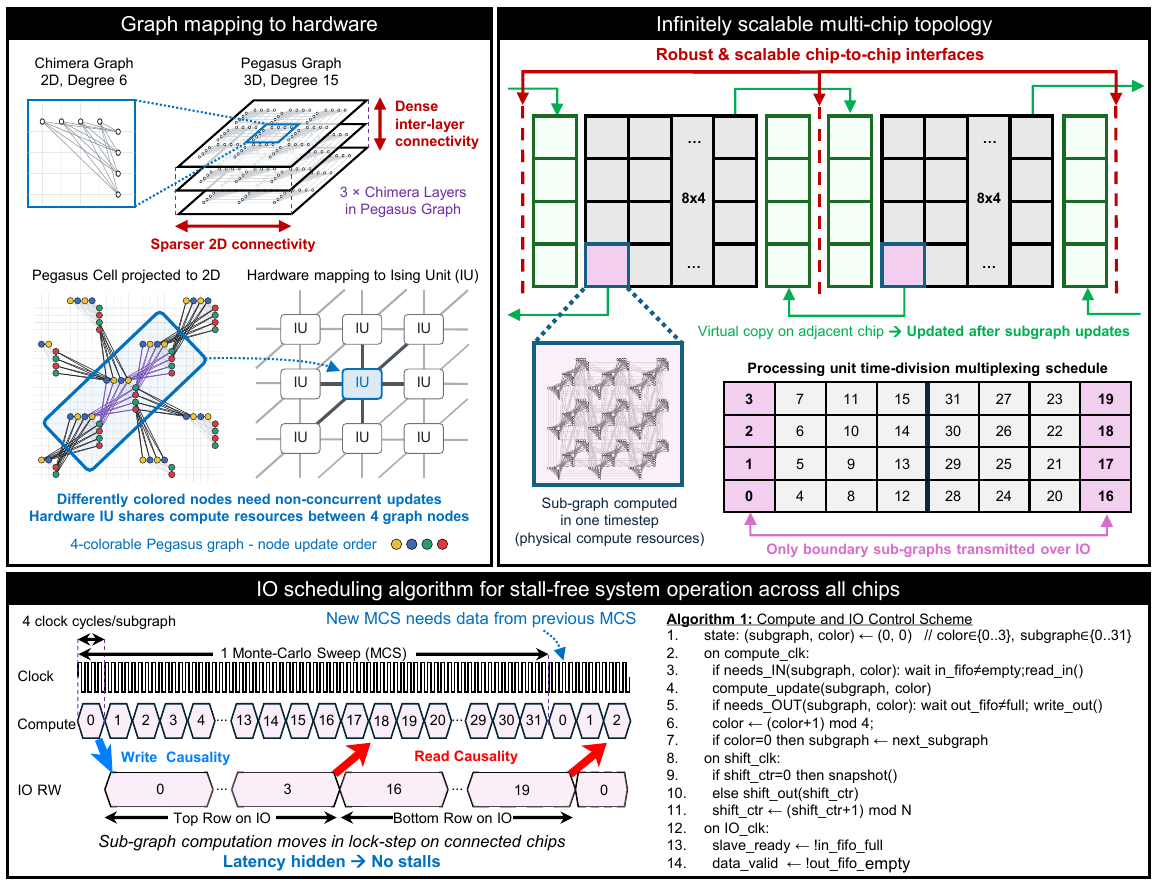}
{Pegasus mapping and four-color update sequence. Boundary-state transfers overlap independent subgraph updates, hiding communication latency when packets arrive before dependent updates. One MCS requires 128 compute cycles without stalls.}
{fig:schedule}
\paperfigure{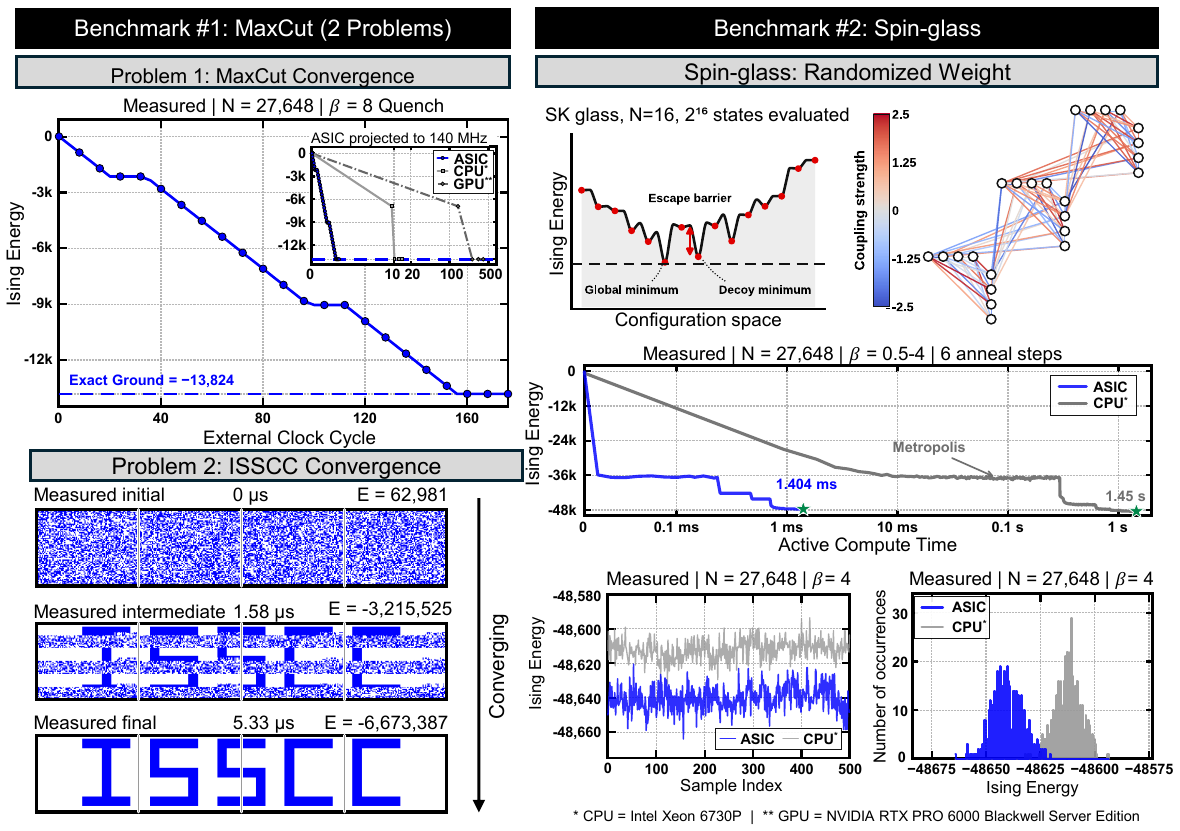}
{Four-chip MaxCut and spin-glass results. Problem 1 shows measured energy versus external clock cycle and an inset projected to 140 MHz. Problem 2 shows convergence of the planted ISSCC image. Spin-glass annealing convergence is marked at 1.404 ms. SK: Sherrington--Kirkpatrick~\cite{sk}; ASIC: application-specific integrated circuit; GPU: graphics processing unit.}
{fig:maxcut}
\paperfigure{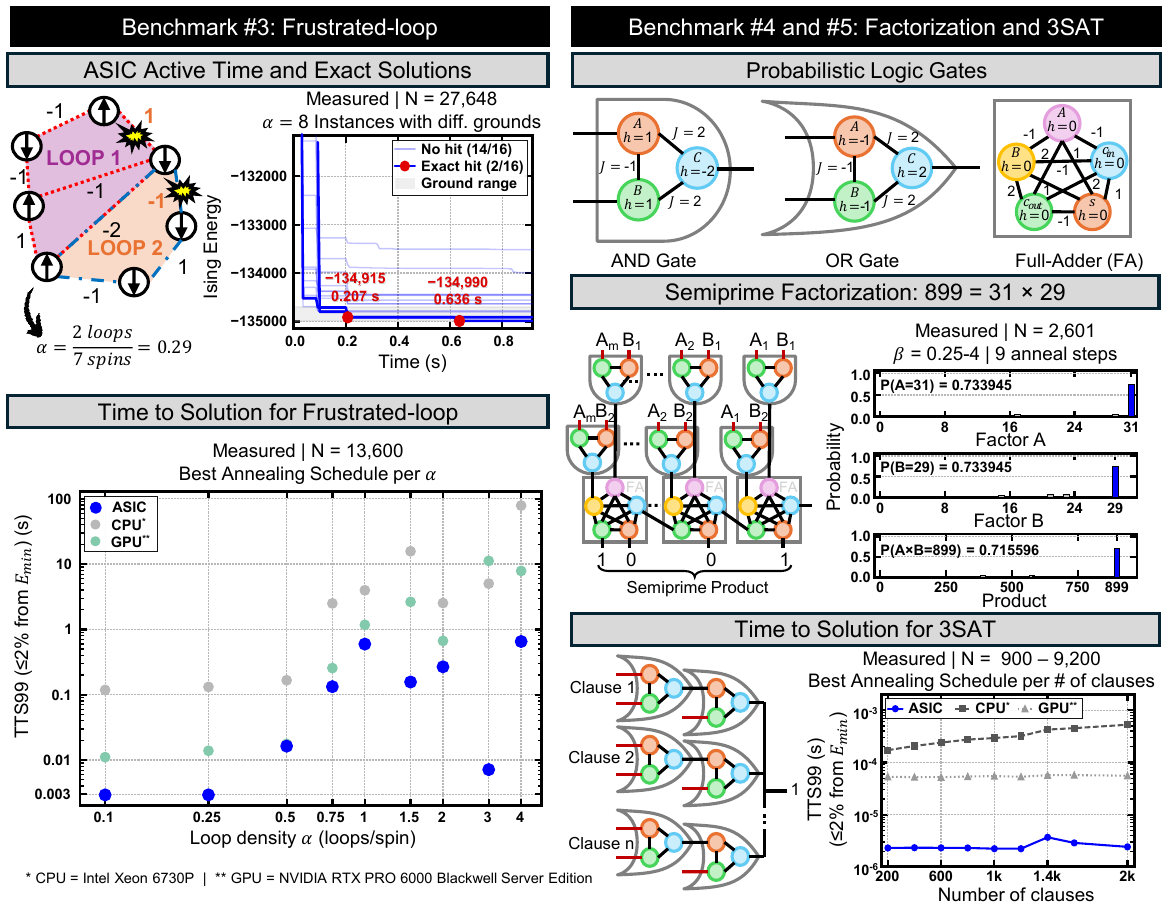}
{Frustrated-loop, factorization, and 3SAT demonstrations. Loop results use 48MHz active time. The $\mathrm{TTS}_{99}$ plots use a 2\% energy-gap target; exact-ground-state hits are marked in the upper-left panel. FA: full adder.}
{fig:benchmarks}
\paperfigure{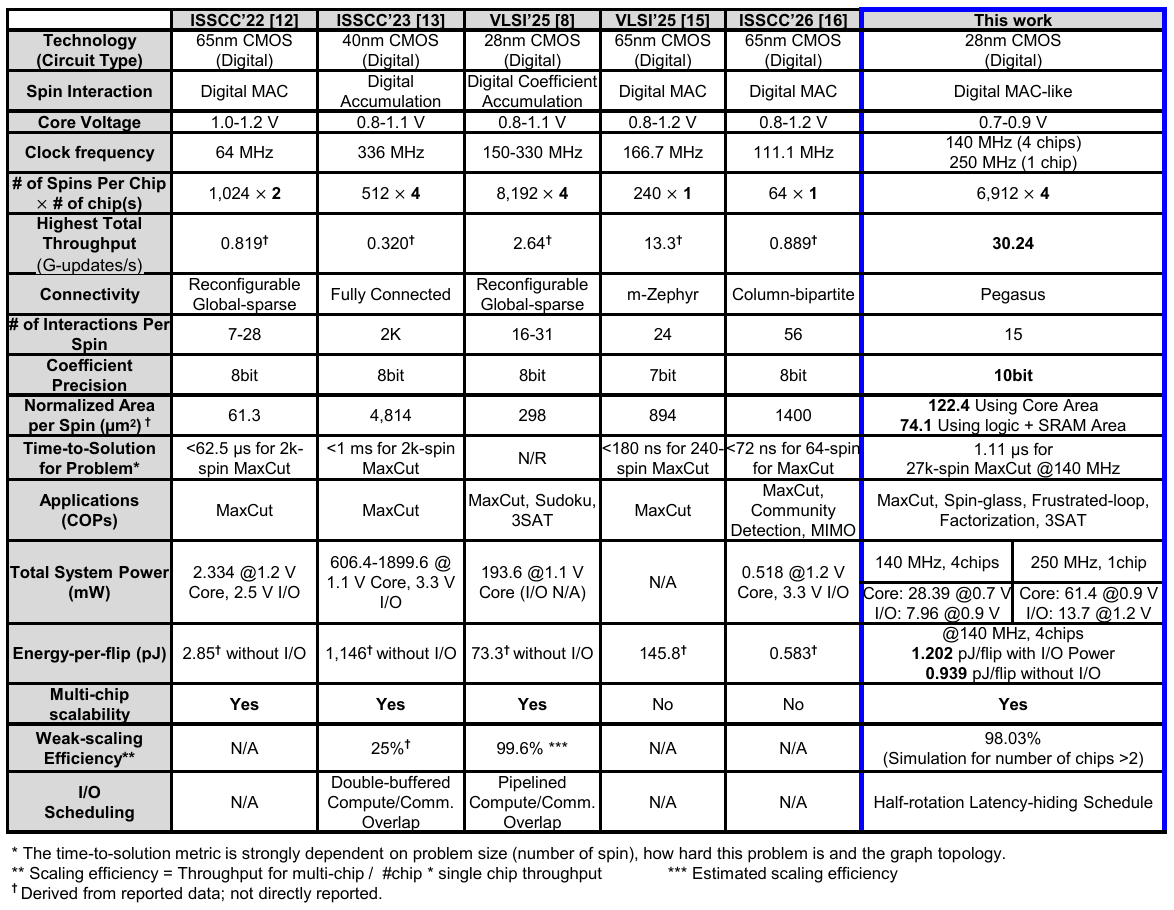}
{Performance summary and comparison with digital Ising accelerators \cite{su,kawamura,kim_sim,mzephyr,cobi}. Core-only energy and SRAM-plus-logic area are compared separately from I/O energy and full core area. Scaling efficiency is simulated. MAC: multiply--accumulate; N/A: not applicable; N/R: not reported.}
{fig:comparison}
\paperfigure{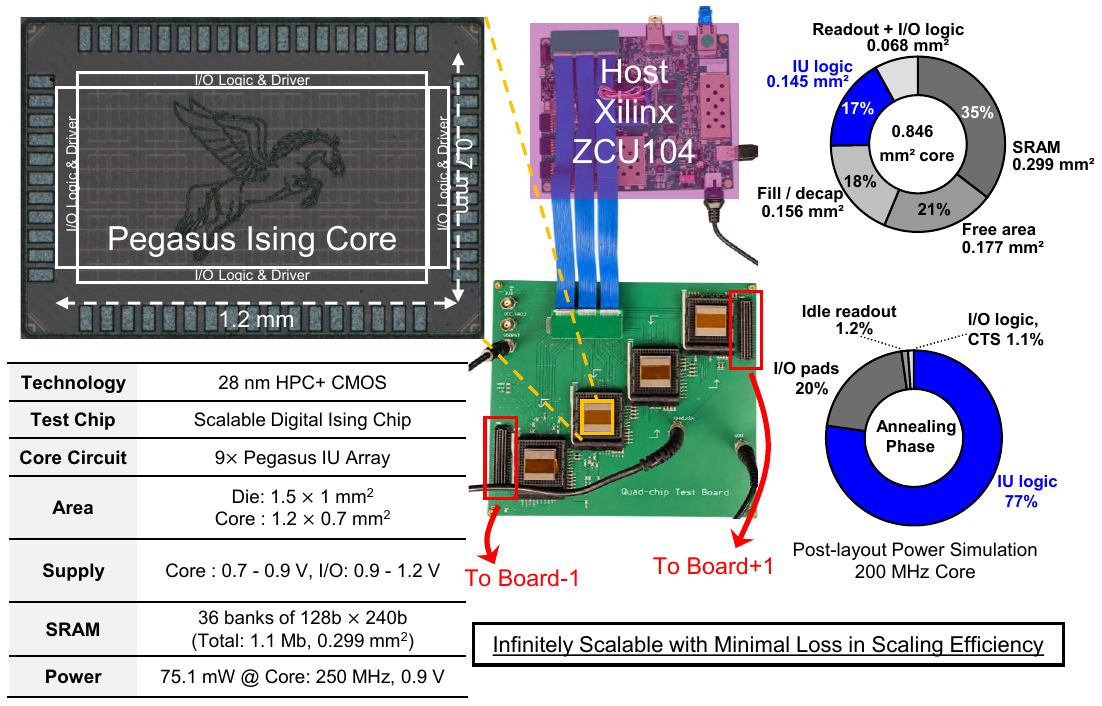}
{Die micrograph, four-chip test platform, area breakdown, and 200MHz post-layout power breakdown during annealing. CTS: clock-tree synthesis.}
{fig:chip}
\clearpage


\clearpage
\begingroup
\small
\begin{thebibliography}{99}
\setlength{\itemsep}{2pt}
\bibitem{lucas} A. Lucas, ``Ising formulations of many NP problems,'' \emph{Frontiers in Physics}, vol. 2, Art. no. 5, 2014.
\bibitem{statica} K. Yamamoto et al., ``STATICA: A 512-spin 0.25M-weight full-digital annealing processor with a near-memory all-spin-updates-at-once architecture for combinatorial optimization with complete spin-spin interactions,'' in \emph{IEEE ISSCC Dig. Tech. Papers}, 2020, pp. 138--140.
\bibitem{dense_sparse} M. M. H. Sajeeb et al., ``Scalable connectivity for Ising machines: Dense to sparse,'' \emph{Physical Review Applied}, vol. 24, no. 1, Art. no. 014005, Jul. 2025.
\bibitem{yamaoka} M. Yamaoka et al., ``20k-spin Ising chip for combinational optimization problem with CMOS annealing,'' in \emph{IEEE ISSCC Dig. Tech. Papers}, 2015, pp. 1--3.
\bibitem{aadit} N. A. Aadit et al., ``Massively parallel probabilistic computing with sparse Ising machines,'' \emph{Nature Electronics}, vol. 5, no. 7, pp. 460--468, Jul. 2022.
\bibitem{takemoto2020} T. Takemoto, M. Hayashi, C. Yoshimura, and M. Yamaoka, ``A $2\times30$k-spin multi-chip scalable CMOS annealing processor based on a processing-in-memory approach for solving large-scale combinatorial optimization problems,'' \emph{IEEE Journal of Solid-State Circuits}, vol. 55, no. 1, pp. 145--156, Jan. 2020.
\bibitem{takemoto2021} T. Takemoto et al., ``A 144Kb annealing system composed of $9\times16$Kb annealing processor chips with scalable chip-to-chip connections for large-scale combinatorial optimization problems,'' in \emph{IEEE ISSCC Dig. Tech. Papers}, 2021, pp. 64--66.
\bibitem{kim_sim} J. Kim and J.-Y. Sim, ``An 8K-spin Ising machine IC with reconfigurable many-body spin interactions and limitless multichip extension,'' in \emph{Proc. Symp. VLSI Technology and Circuits}, 2025, pp. 1--3. \href{https://doi.org/10.23919/VLSITechnologyandCir65189.2025.11074989}{doi:10.23919/VLSITechnologyandCir65189.2025.11074989}.
\bibitem{pegasus} N. Dattani, S. Szalay, and N. Chancellor, ``Pegasus: The second connectivity graph for large-scale quantum annealing hardware,'' arXiv:1901.07636, 2019.
\bibitem{niazi} S. Niazi et al., ``Training deep Boltzmann networks with sparse Ising machines,'' \emph{Nature Electronics}, vol. 7, no. 7, pp. 610--619, Jul. 2024.
\bibitem{bae} J. Bae, C. Shim, and B. Kim, ``e-Chimera: A scalable SRAM-based Ising macro with enhanced-Chimera topology for solving combinatorial optimization problems within memory,'' in \emph{IEEE ISSCC Dig. Tech. Papers}, 2024, pp. 286--288.
\bibitem{su} Y. Su, T. T.-H. Kim, and B. Kim, ``FlexSpin: A scalable CMOS Ising machine with 256 flexible spin processing elements for solving complex combinatorial optimization problems,'' in \emph{IEEE ISSCC Dig. Tech. Papers}, 2022, pp. 274--276. \href{https://doi.org/10.1109/ISSCC42614.2022.9731680}{doi:10.1109/ISSCC42614.2022.9731680}.
\bibitem{kawamura} K. Kawamura et al., ``Amorphica: 4-replica 512 fully connected spin 336MHz metamorphic annealer with programmable optimization strategy and compressed-spin-transfer multi-chip extension,'' in \emph{IEEE ISSCC Dig. Tech. Papers}, 2023, pp. 42--44. \href{https://doi.org/10.1109/ISSCC42615.2023.10067504}{doi:10.1109/ISSCC42615.2023.10067504}.
\bibitem{chu} Y.-C. Chu, Y.-C. Lin, Y.-C. Lo, and C.-H. Yang, ``A fully integrated annealing processor for large-scale autonomous navigation optimization,'' in \emph{IEEE ISSCC Dig. Tech. Papers}, 2024, pp. 488--490. \href{https://doi.org/10.1109/ISSCC49657.2024.10454294}{doi:10.1109/ISSCC49657.2024.10454294}.
\bibitem{mzephyr} Y. Wu, J. Bae, C. Shim, and B. Kim, ``m-Zephyr: A digital in-memory Ising chip with 240 spins featuring enhanced connectivity based on a modified 3D Zephyr topology,'' in \emph{Proc. Symp. VLSI Technology and Circuits}, 2025, pp. 1--3.
\bibitem{cobi} Y. Wu, J. Bae, S. Shin, and B. Kim, ``COBI: A degree-of-56 column-bipartite densely connected digital Ising chip with 8b spin coefficients,'' in \emph{IEEE ISSCC Dig. Tech. Papers}, 2026, pp. 182--184. \href{https://doi.org/10.1109/ISSCC49663.2026.11409141}{doi:10.1109/ISSCC49663.2026.11409141}.
\bibitem{choi} V. Choi, ``Minor-embedding in adiabatic quantum computation: I. The parameter setting problem,'' \emph{Quantum Information Processing}, vol. 7, no. 5, pp. 193--209, Oct. 2008.
\bibitem{minorminer} J. Cai, W. G. Macready, and A. Roy, ``A practical heuristic for finding graph minors,'' arXiv:1406.2741, 2014.
\bibitem{bunyk} P. I. Bunyk et al., ``Architectural considerations in the design of a superconducting quantum annealing processor,'' \emph{IEEE Transactions on Applied Superconductivity}, vol. 24, no. 4, Art. no. 1700110, Aug. 2014.
\bibitem{camsari} K. Y. Camsari, R. Faria, B. M. Sutton, and S. Datta, ``Stochastic p-bits for invertible logic,'' \emph{Physical Review X}, vol. 7, no. 3, Art. no. 031014, Jul. 2017.
\bibitem{metropolis} N. Metropolis, A. W. Rosenbluth, M. N. Rosenbluth, A. H. Teller, and E. Teller, ``Equation of state calculations by fast computing machines,'' \emph{Journal of Chemical Physics}, vol. 21, no. 6, pp. 1087--1092, Jun. 1953.
\bibitem{kirkpatrick} S. Kirkpatrick, C. D. Gelatt, and M. P. Vecchi, ``Optimization by simulated annealing,'' \emph{Science}, vol. 220, no. 4598, pp. 671--680, May 1983.
\bibitem{hen} I. Hen, J. Job, T. Albash, T. F. R{\o}nnow, M. Troyer, and D. A. Lidar, ``Probing for quantum speedup in spin-glass problems with planted solutions,'' \emph{Physical Review A}, vol. 92, no. 4, Art. no. 042325, Oct. 2015.
\bibitem{ronnow} T. F. R{\o}nnow et al., ``Defining and detecting quantum speedup,'' \emph{Science}, vol. 345, no. 6195, pp. 420--424, Jul. 2014.
\bibitem{sk} D. Sherrington and S. Kirkpatrick, ``Solvable model of a spin-glass,'' \emph{Physical Review Letters}, vol. 35, no. 26, pp. 1792--1796, Dec. 1975.
\end{thebibliography}

\endgroup
\end{document}